\documentclass[11pt]{article}

\usepackage[T1]{fontenc}
\usepackage{lmodern}
\usepackage[letterpaper,margin=1.15in]{geometry}
\usepackage{graphicx}
\usepackage{booktabs}
\usepackage{amsmath}
\usepackage{amssymb}
\usepackage{siunitx}
\usepackage[table]{xcolor}
\usepackage{caption}
\usepackage{microtype}
\usepackage[hidelinks]{hyperref}
\usepackage[round,authoryear]{natbib}
\usepackage{fancyhdr}

\hypersetup{
  pdftitle={Design Creativity Bench: Measuring creativity in LLM-Generated UI},
  pdfauthor={Aman Rusia, Abhijit Bhole, Prashank Gupta, Dipanjan Dey},
}

\fancypagestyle{preprint}{\fancyhf{}\fancyhead[L]{\small\itshape Preprint.}\fancyfoot[C]{\thepage}}
\newcommand{\dbench}{\mathcal{D}_{B}}
\newcommand{\dproj}{\mathcal{D}_{P}}
\newcommand{\dladder}{\mathcal{D}_{L}}
\newcommand{\dhuman}{\mathcal{D}_{B}^{H}}
\newcommand{\dcalib}{\mathcal{D}_{C}}
\newcommand{\dptrain}{\mathcal{D}_{P}^{\mathrm{train}}}
\newcommand{\dpval}{\mathcal{D}_{P}^{\mathrm{val}}}

\title{\bfseries Design Creativity Bench: Measuring creativity in LLM-Generated UI}
\author{Aman Rusia\thanks{Correspondence to \href{mailto:aman@kombai.com}{aman@kombai.com}.} \quad Abhijit Bhole \quad Prashank Gupta \quad Dipanjan Dey\\[2pt]
\small Kombai Inc. \quad \url{https://kombai.com}}
\date{}

\begin{document}
\maketitle
\thispagestyle{preprint}

\begin{abstract}
\noindent
As leading LLMs improve on capability evaluations, their limitations in producing creative outputs on design tasks remain insufficiently characterised. Our work introduces Design Creativity Bench, a benchmark that evaluates diversity and appropriateness in UI designs. It measures distinctiveness among models on the same prompt (originality), how much a model's designs change between two prompts for the same UI goal in different product domains (creative range), and the share of a brief's acceptance criteria each design meets (appropriateness). Originality is 0.592 for same-prompt design pairs from different models (95\% CI [0.582, 0.602]), far below the 0.764 for same-prompt human--model pairs (95\% CI [0.751, 0.778]). Creative range is 0.581 across models (95\% CI [0.567, 0.597]), against 0.902 for human designs (95\% CI [0.884, 0.919]). Appropriateness is above 90\% for every model in our test set, and the best model reaches 99.2\%, slightly above the 98.0\% for human designs. Our work shows that the default output of LLMs, though generally appropriate, is substantially more repetitive than the human baseline. This calls for strong measures to address the issue. A live leaderboard with newer models is maintained at \url{https://kombai.com/benchmark/}.

\end{abstract}

\begin{figure}[t]
\centering
\includegraphics[width=\textwidth]{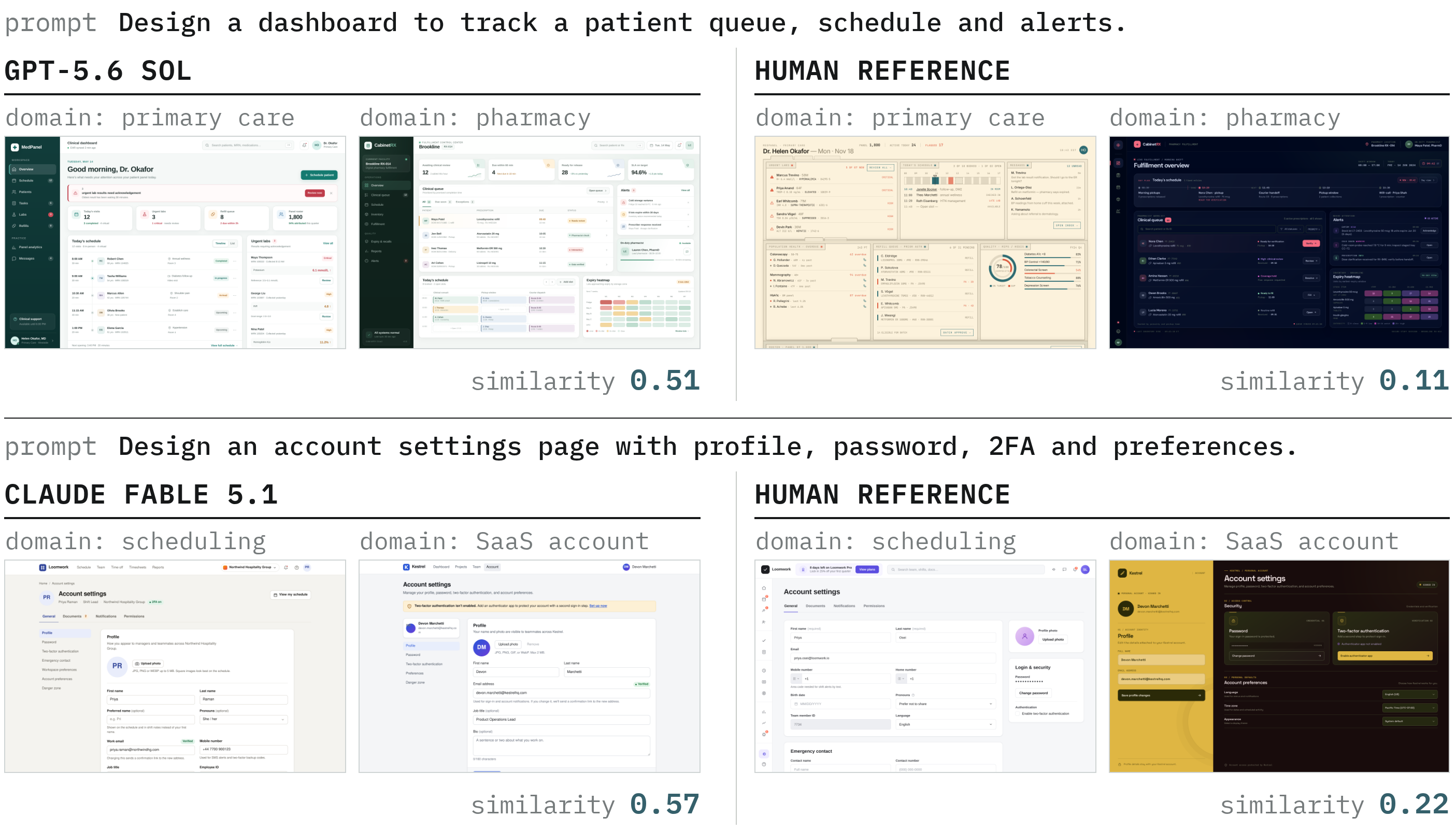}
\caption{\textbf{Diversity in LLM UI generation compared with human designs.} The examples are drawn from the benchmark $\dbench$. The model-generated designs are markedly more similar across domain and content variants than the corresponding human reference designs. Each row pairs two prompts that share the goal shown (paraphrased) and differ in product domain and content; similarity is the pair score, from 0 to 1, where higher means more alike. We ask models to generate desktop designs for a $1280\times720$ target viewport.}
\label{fig:hero}
\end{figure}

\section{Introduction}
\label{sec:intro}

Creativity and originality are important qualities in user interface (UI) design \citep{lavie2004aesthetics}. While LLM capabilities appear to be on an exponential growth trajectory \citep{kwa2025horizon,xiao2025densing}, the creativity of LLMs in UI design remains largely unknown.

Previous studies have reported poorer creativity in LLMs compared with humans on text and code generation tasks \citep{chakrabarty2024artifice,lu2025creativity,wenger2025homogeneity}. The dominant account attributes output homogeneity to post-training alignment
\citep{kirk2024rlhf,padmakumar2024writing,peeperkorn2024temperature,jiang2025hivemind,west2025aligned,shypula2025effective}, but recent work argues that it may originate in the likelihood-maximisation objective itself \citep{fortier2026convergence}, suggesting that post-alignment interventions alone may not resolve the problem.

Creativity is frequently characterized in terms of appropriateness and novelty \citep{guilford1950creativity,runco2012standard,peeperkorn2024temperature}. Studies of LLMs on open-ended tasks commonly assess novelty through output diversity, i.e.\ how different outputs are from one another \citep{tevet2021diversity,zhang2025novelty,jiang2025hivemind,wenger2025homogeneity,padmakumar2024writing}. Several studies have examined whether simple prompting and sampling changes can increase this diversity, including prompt paraphrasing, creativity-oriented system prompts, in-context regeneration, temperature variation, and alternative sampling schemes \citep{shur2026growing,zhang2025novelty,jiang2025hivemind,peeperkorn2024temperature,wenger2025homogeneity}. These changes often have limited effects \citep{peeperkorn2024temperature,jiang2025hivemind,wenger2025homogeneity,zhang2025novelty}, and even where larger gains occur, independently generated outputs from the individual models evaluated in these studies remain below human baselines \citep{shur2026growing,zhang2025novelty}.

These studies identify persistent limitations in the diversity of generated outputs, even when individual responses are otherwise appropriate. However, findings from text and code generation tasks do not establish the extent of this problem in UI design tasks. This gap motivates us to examine whether LLM-generated UI designs exhibit similar limitations.

To this end, we introduce Design Creativity Bench.\footnote{Results for models released after this paper, evaluated with the same protocol, are maintained at \url{https://kombai.com/benchmark/}.} This benchmark uses pairs of open-ended design briefs that share a design goal and UI archetype but differ in product domain and content requirements. Our open-ended UI briefs leave many design decisions unspecified, permitting multiple plausible solutions; Appendix~\ref{app:entropy} illustrates these choices for one brief. Changes in domain and content provide additional reasons to vary those decisions. We evaluate the HTML outputs generated by LLMs for these briefs using three complementary measures.
\begin{itemize}
\item \textbf{Originality} measures differences between models on the same prompt, reported as mean dissimilarity of those design pairs.
\item \textbf{Creative range} measures variation within a model across the paired briefs, reported as the mean distinctiveness of its design pairs across the two briefs.
\item \textbf{Appropriateness} measures percentage of a brief’s acceptance criteria that a design satisfies.

\end{itemize}

We also combine the three measures into an overall score, a summary of each model's creativity, and report it along with the individual measures.

We also collect human-created designs under the same briefs and evaluate them using the same metrics. These human-created designs form our human gold baseline, an empirical reference for design creativity as reflected in variation among acceptable designs. Throughout the paper, we refer to this set as the human baseline or human reference.

Together, these measures let us evaluate UI design creativity as reflected in variation among acceptable designs, using the human baseline as a point of comparison.

\section{Details of the metrics}
\label{sec:measures}

\subsection{Measuring how much two designs repeat}
\label{sec:embedding}

\begin{figure}[t]
\centering
\includegraphics[width=\textwidth]{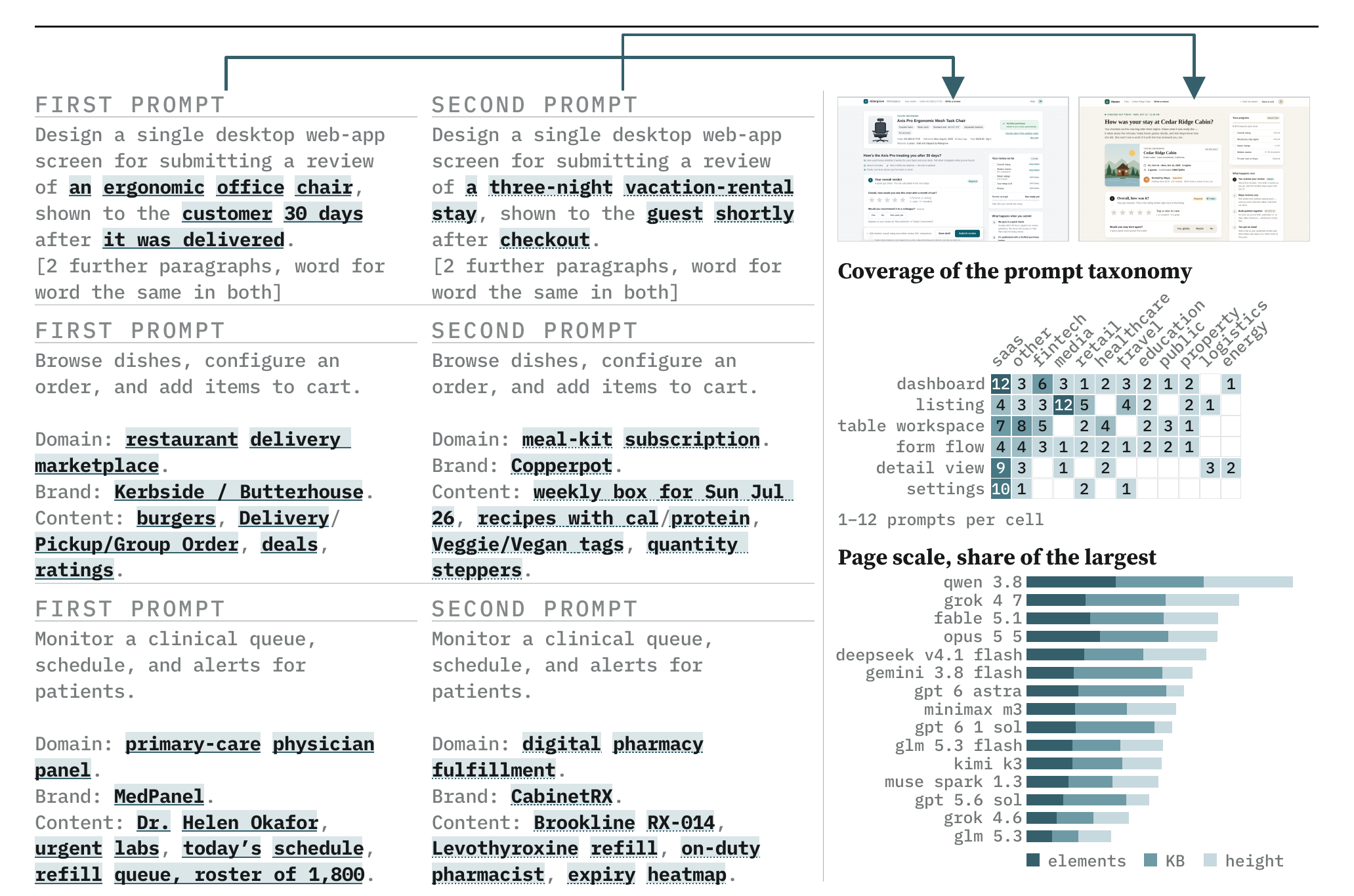}
\caption{\textbf{Prompt pair examples to measure creative range.} Prompt inventory for $\dbench$. Left: three
prompt pairs. The two prompts in each pair share the goal but differ in domain and content.
Top right: what Claude Fable 5.1 returned for the first pair, each design linked by an arrow to the prompt
that produced it. Right: the taxonomy covered by the 160 prompts, with screen
archetype plotted against product vertical and prompt counts shown using a single hue because each cell value is a
magnitude; and how much page content each model generates.}
\label{fig:corpus}
\end{figure}

We first select a method to obtain a similarity score $S$ for a pair; the distinctiveness of the pair is $1-S$. To check the score against human judgement, we labeled 100 design pairs as distinct, not distinct, or unsure (no pair was labelled unsure). On this dataset ($\dcalib$), the selected similarity score separates the two classes with an AUC of $0.85$. $\dcalib$ was not used to select or tune the scorer, so this is an out-of-sample check. Appendix~\ref{app:threshold} provides details.

The similarity-score method itself was selected from various options based on human-agreement and robustness evaluations. We describe it next.

\paragraph{Embedding and scoring.}
Comparing UI designs is more involved than comparing image or text similarity because more
invariants are involved: a product card with a different title, icon and spacing is materially the
same design element.

Other visual variants that do not change the semantics of a UI element include:
\begin{itemize}
\item Viewport width changes
\item Responsive reflow
\item Text changes
\item Changes in graphics or icons
\item Spacing changes (up to some limit)
\item Font-scale changes
\item Size of the dynamically sized group ($N$ rows of tables, $N$ cards in a grid)
\end{itemize}

At the same time, some changes do alter the design: deleting or duplicating child elements, reordering them, changing the layout direction, or flattening and reparenting parts of the hierarchy. We call these structural edits. Raw image embeddings treat them almost like the invariant changes above, so a score built on them cannot tell a new layout from a restyled one. We show that embeddings trained to ignore the invariant changes and react to the structural ones have higher agreement with human labels. 

We use both kinds of edits to train a linear layer over image embeddings (loosely termed a projection layer, as it reduces the vector to 64 dimensions) on a separate dataset, $\dproj$. The dataset $\dproj$ and loss functions are outlined in Appendix~\ref{app:projection}. We experimented with leading image embeddings (Gemini Embedding \citep{googleembedding}) and specialized UI embeddings (UIClip \citep{wu2024uiclip}, a CLIP model \citep{radford2021learning} fine-tuned on interface screenshots). We observed that UIClip has lower loss on larger (page-level) segments, while Gemini performs better on smaller segments; therefore, we also experimented with a weighted, rescaled combination of the two embeddings, henceforth called the \textbf{hybrid embedding}.

We also observed that pages are often large and that full-page image embeddings lose finer differences. Our work obtains page segments and compares pages based on segment-level differences. Segment-level similarity is obtained by first segmenting the whole page into containers (top-level sections, compound components, stacks, etc.), embedding each segment's rendered crop, matching segments between pairs (allowing null matches), assigning an importance weight to each segment pair, and then summing the weighted scores (Appendix~\ref{app:segments} defines a segment and how each one is captured). We treat the design's color system as a separate top-level comparison; to avoid double counting, we train the embedding layers to be invariant to color systems (details in Appendix~\ref{app:colour}).

In addition to embedding-based approaches, we also consider LLM-as-judge methods, which receive a design pair and assign a similarity score. In Figure~\ref{fig:human-agreement}, the scoring approaches are evaluated against a comparison dataset, $\dhuman$, labeled by three human reviewers (Appendix~\ref{app:human}). While our fine-tuned, hybrid, segment-wise scoring outperformed whole-page embeddings and raw embeddings, LLM-as-judge had the highest agreement with human consensus. On the other hand, the best LLM-as-judge (Grok 4.6), when used to obtain a single similarity score, had robustness issues on our design-perturbation dataset $\dladder$. Its score was not monotonic under graded perturbations, as shown in Figure~\ref{fig:ladders}. We choose the best embedding-based approach as the score for our leaderboard because it passes all monotonicity checks.

\subsection{Measuring originality}
\label{sec:distinctiveness}

For each model, we compare its designs with those of every other model designing against
the same brief. Models from the same lab are left out from the pairings. Originality is the mean
of $1-S$ over these pairs, where $S$ is the pair similarity score. Higher values indicate greater distinctiveness from the model population.
The human reference uses the same measure on human--model pairs. 

\subsection{Measuring creative range}
\label{sec:creativity}

Creative range applies the same pair score within one model. Each task
is issued as two prompts that keep the goal and UI archetype but change the product domain and
content description (Figure~\ref{fig:corpus}). For each model and task, the model's two designs
form one pair. A model's creative range is the mean of $1-S$ over its two-prompt pairs.
Higher values indicate that the model's
designs respond to the change in the brief. The human reference is scored the same way on the
two human designs of each task.

\begin{table}[t]
\centering
\small
\caption{Participating models, with the reasoning effort each model was generated with. Every model
runs at its highest effort except Claude Opus 5.5$^\dagger$ which runs at xhigh since at max it runs out of the time budget.}
\label{tab:models}
\begin{tabular}{@{}ll@{}}
\toprule
\textbf{Model (effort)} & \textbf{Provider} \\
\midrule
Qwen3.8 Max (max)                  & Alibaba   \\
Claude Fable 5.1 (max)             & Anthropic \\
Claude Opus 5.5 (xhigh)$^\dagger$  & Anthropic \\
DeepSeek V4.1 Flash (max)          & DeepSeek  \\
Gemini 3.8 Flash (high)            & Google    \\
Muse Spark 1.3 (xhigh)             & Meta      \\
MiniMax M3 (high)                  & MiniMax   \\
Kimi K3 (max)                      & Moonshot  \\
GPT-5.6 Sol (max)                  & OpenAI    \\
GPT-6 Astra (max)                  & OpenAI    \\
GPT-6.1 Sol (max)                  & OpenAI    \\
Grok 4.6 (xhigh)                   & xAI       \\
Grok 4.7 (xhigh)                   & xAI       \\
GLM-5.3 (max)                      & Z.ai      \\
GLM-5.3-Flash (max)                & Z.ai      \\
\midrule
Human gold baseline                & ---       \\
\bottomrule
\end{tabular}
\end{table}

\subsection{Measuring appropriateness}
For each design brief, we first define a set of yes/no criteria that together specify what an appropriate and correct design must satisfy: the requested screen, the content and values the brief states, controls for the actions it asks for, and the correctness of any values derived from them. Each criterion is also tagged as critical when failing it means the design does not do the job the brief asks, such as a wrong page type, a missing or wrong required value, or no control for a core action. We then judge each generated design against its brief's criteria and report the percentage of criteria met across a model's designs. Appendix~\ref{app:appropriateness} gives an example criteria set.

\section{Results}
\label{sec:results}

Fifteen leading models (Table~\ref{tab:models}) were used to generate 160 designs each (2,400 primary model designs in total) across 80 tasks and 80 differentiated pairs per model ($\dbench$; Appendix~\ref{app:prompts} gives the breakdown). Originality and creative range were measured using an embedding method with high human agreement on two different evaluations (AUC $0.87$ on $\dhuman$, AUC $0.85$ on $\dcalib$). Human designers received the benchmark briefs and access to public inspiration libraries, such as Behance and Kombai Gallery. They created designs for these briefs by adapting publicly available design references to the brief requirements. Reference access was intended to support high-quality design within time constraints and reflect how designers often use inspirations in practice.

\paragraph{Model-generated designs resemble one another more than they resemble the human reference.}
Table~\ref{tab:diversity} ranks the models by originality. We take every pair of designs from two different models designing from the same prompt and average their distinctiveness $1-S$. Across the fifteen models, originality is 0.592 (95\% CI [0.582, 0.602]) over 15,840 unique model--model pairs, compared with 0.764 [0.751, 0.778] for 2,400 human--model pairs.
 The pooled interval uses 20,000 bootstrap resamples of whole tasks \citep{efron1979bootstrap,field2007bootstrapping} (seed 7), retaining every pair within each sampled task and recomputing the pooled mean; it conditions on the observed models, generations and fitted scorer. The most original models are GPT-6 Astra (0.646) and Grok 4.7 (0.643), and the least original, Kimi K3, scores 0.547. Even the most original model is 0.12 below the human row, and no model interval reaches it.

\begin{table}[htbp]
\centering
\small
\caption{Originality. Two models from different labs design against the same prompt. We report
the mean distinctiveness $1-S$ of those pairs. A high value means a
model differs from the others. Each model contributes 160 pairs per partner from
another lab: 1,920 to 2,240 pairs. The human row compares each human design with model designs from
all 15 models. Intervals
are task-blocked bootstraps. Bold marks the highest model value. Cost and time are medians per
generated design.}
\label{tab:diversity}
\begin{tabular}{@{}rlrrrr@{}}
\toprule
Rank & Model & Originality & 95\% CI & Cost (USD) & Time (min) \\
\midrule
1 & GPT-6 Astra & \textbf{0.646} & [0.631, 0.661] & 2.564 & 13.8 \\
2 & Grok 4.7 & 0.643 & [0.631, 0.655] & 0.593 & 18.8 \\
3 & GPT-6.1 Sol & 0.642 & [0.628, 0.657] & 0.460 & 15.8 \\
4 & Gemini 3.8 Flash & 0.637 & [0.621, 0.653] & 0.123 & 0.9 \\
5 & Qwen3.8 Max & 0.623 & [0.609, 0.636] & 0.309 & 18.0 \\
6 & Grok 4.6 & 0.591 & [0.579, 0.602] & 0.100 & 3.5 \\
7 & GPT-5.6 Sol & 0.587 & [0.574, 0.600] & 0.598 & 4.4 \\
8 & MiniMax M3 & 0.587 & [0.575, 0.599] & 0.024 & 2.5 \\
9 & GLM-5.3 & 0.579 & [0.567, 0.590] & 0.043 & 2.5 \\
10 & DeepSeek V4.1 Flash & 0.576 & [0.565, 0.588] & 0.051 & 5.3 \\
11 & Muse Spark 1.3 & 0.568 & [0.558, 0.578] & 0.080 & 1.1 \\
12 & Claude Fable 5.1 & 0.558 & [0.547, 0.570] & 5.329 & 13.9 \\
13 & Claude Opus 5.5 & 0.557 & [0.548, 0.567] & 1.557 & 13.1 \\
14 & GLM-5.3-Flash & 0.553 & [0.543, 0.564] & 0.022 & 10.0 \\
15 & Kimi K3 & 0.547 & [0.536, 0.558] & 0.570 & 17.0 \\
\midrule
--- & Human reference & 0.764 & [0.751, 0.778] & --- & --- \\
\bottomrule
\end{tabular}
\end{table}

\paragraph{Differentiating the prompt does not differentiate the design.}
Table~\ref{tab:creativity} ranks the models on creative range. The model is fixed and the prompt
changes: both prompts state the same goal in the same page archetype and differ only in product
domain and sample content. Across the fifteen models, the mean distinctiveness of those pairs is
0.581 (95\% CI [0.567, 0.597]), against 0.902 [0.884, 0.919] for the human pairs built the same way.
The pooled interval resamples whole tasks as in Table~\ref{tab:diversity} (20,000 resamples, seed 7).
The spread is wide -- GLM-5.3 scores 0.634 and GPT-6 Astra 0.472 -- but no model interval reaches
the human value. The least distinct archetype is settings (0.554) and the most distinct is listing
(0.590).

\begin{figure}[t]
\centering
\includegraphics[width=\textwidth]{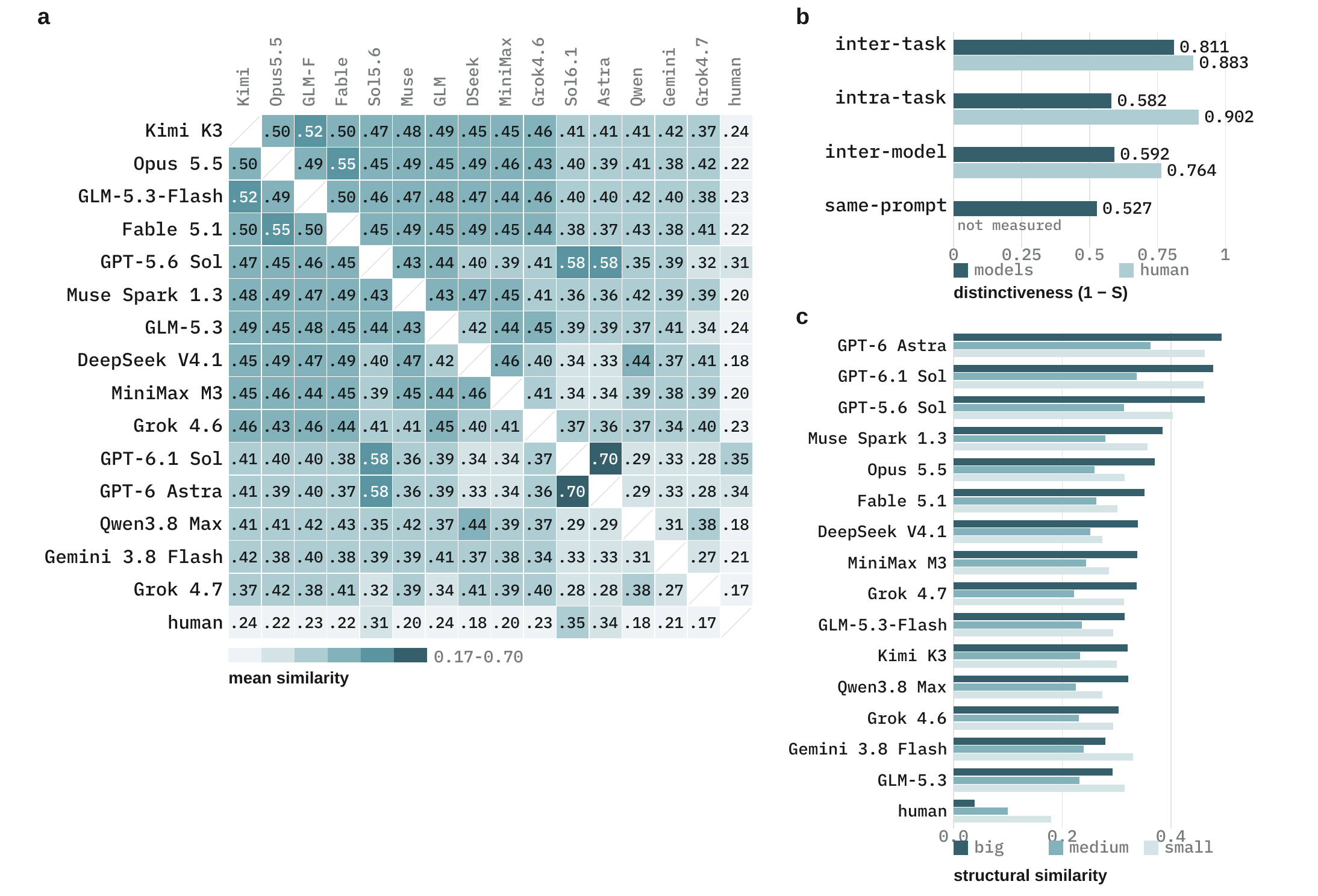}
\caption{\textbf{Design distinctiveness across comparison conditions.}
Results on $\dbench$.
(a) Pairwise similarity for different models on same prompts; human cells compare a model
with a human reference. Higher score is worse. (b) Distinctiveness score on four different comparisons: different tasks; differentiated prompts within one
task; different models on same prompt; and independent generations of an identical prompt.
The first two human bars compare two human designs; the inter-model human bar compares models
with humans. Higher is better. (c) Segment wise similarity grouped by segment size: small ($<2\%$), medium ($2$--$10\%$), and
big ($\geq10\%$), where percentages are with respect to view port size. Higher is worse.}
\label{fig:inter-model}
\end{figure}

What makes this an important result is that there are enough changes in the two prompts to motivate diversity. Indeed, when
the same model designs two unrelated tasks, the mean distinctiveness is 0.811
(Figure~\ref{fig:inter-model}b).

\begin{table}[htbp]
\centering\small
\caption{Creative range. Each task is issued as two prompts that share a goal and page archetype but differ in product domain and content. Cells are the mean distinctiveness $1-S$ of a model's two designs for the two prompts, split by archetype; higher means the design changed more with the brief. Each model contributes 80 pairs, one per task: Listing 18, Dashboard 18, Form flow 11, Table 16, Detail 10, Settings 7. The overall column pools all 80 pairs, and its interval is a task-blocked bootstrap. The human row pairs the two human designs of each task. Rows are ranked on the overall value and bold marks the highest model value in each column; the human reference is excluded from that comparison.}
\label{tab:creativity}
\footnotesize\setlength{\tabcolsep}{3.6pt}
\begin{tabular}{@{}rlrrrrrrrr@{}}
\toprule
Rank & Model & Listing & Dashboard & Form flow & Table & Detail & Settings & Overall & 95\% CI \\
\midrule
1 & GLM-5.3 & 0.63 & 0.63 & 0.68 & \textbf{0.65} & 0.61 & 0.59 & \textbf{0.634} & [0.607, 0.662] \\
2 & Gemini 3.8 Flash & 0.63 & 0.62 & \textbf{0.68} & 0.63 & 0.60 & 0.58 & 0.628 & [0.602, 0.652] \\
3 & Grok 4.6 & 0.63 & \textbf{0.65} & 0.64 & 0.61 & 0.61 & 0.54 & 0.625 & [0.599, 0.650] \\
4 & Qwen3.8 Max & \textbf{0.65} & 0.62 & 0.60 & 0.60 & 0.62 & 0.59 & 0.617 & [0.597, 0.639] \\
5 & Kimi K3 & 0.60 & 0.59 & 0.62 & 0.63 & \textbf{0.67} & \textbf{0.62} & 0.615 & [0.593, 0.636] \\
6 & GLM-5.3-Flash & 0.60 & 0.63 & 0.58 & 0.63 & 0.62 & 0.60 & 0.611 & [0.590, 0.632] \\
7 & Grok 4.7 & 0.62 & 0.63 & 0.58 & 0.58 & 0.59 & 0.60 & 0.603 & [0.583, 0.624] \\
8 & MiniMax M3 & 0.60 & 0.61 & 0.59 & 0.62 & 0.55 & 0.62 & 0.602 & [0.580, 0.624] \\
9 & DeepSeek V4.1 Flash & 0.57 & 0.61 & 0.64 & 0.57 & 0.61 & 0.59 & 0.596 & [0.576, 0.617] \\
10 & Claude Fable 5.1 & 0.62 & 0.58 & 0.56 & 0.60 & 0.56 & 0.57 & 0.589 & [0.562, 0.615] \\
11 & Claude Opus 5.5 & 0.57 & 0.58 & 0.56 & 0.59 & 0.57 & 0.54 & 0.574 & [0.551, 0.597] \\
12 & Muse Spark 1.3 & 0.57 & 0.57 & 0.58 & 0.56 & 0.56 & 0.51 & 0.565 & [0.545, 0.585] \\
13 & GPT-5.6 Sol & 0.55 & 0.48 & 0.50 & 0.51 & 0.51 & 0.43 & 0.505 & [0.482, 0.527] \\
14 & GPT-6.1 Sol & 0.50 & 0.49 & 0.45 & 0.51 & 0.47 & 0.47 & 0.487 & [0.462, 0.513] \\
15 & GPT-6 Astra & 0.50 & 0.46 & 0.46 & 0.48 & 0.48 & 0.43 & 0.472 & [0.448, 0.498] \\
\midrule
--- & Human reference & 0.89 & 0.91 & 0.91 & 0.93 & 0.86 & 0.89 & 0.902 & [0.884, 0.919] \\
\bottomrule
\end{tabular}
\end{table}

\begin{figure}[t]
\centering
\includegraphics[width=0.72\textwidth]{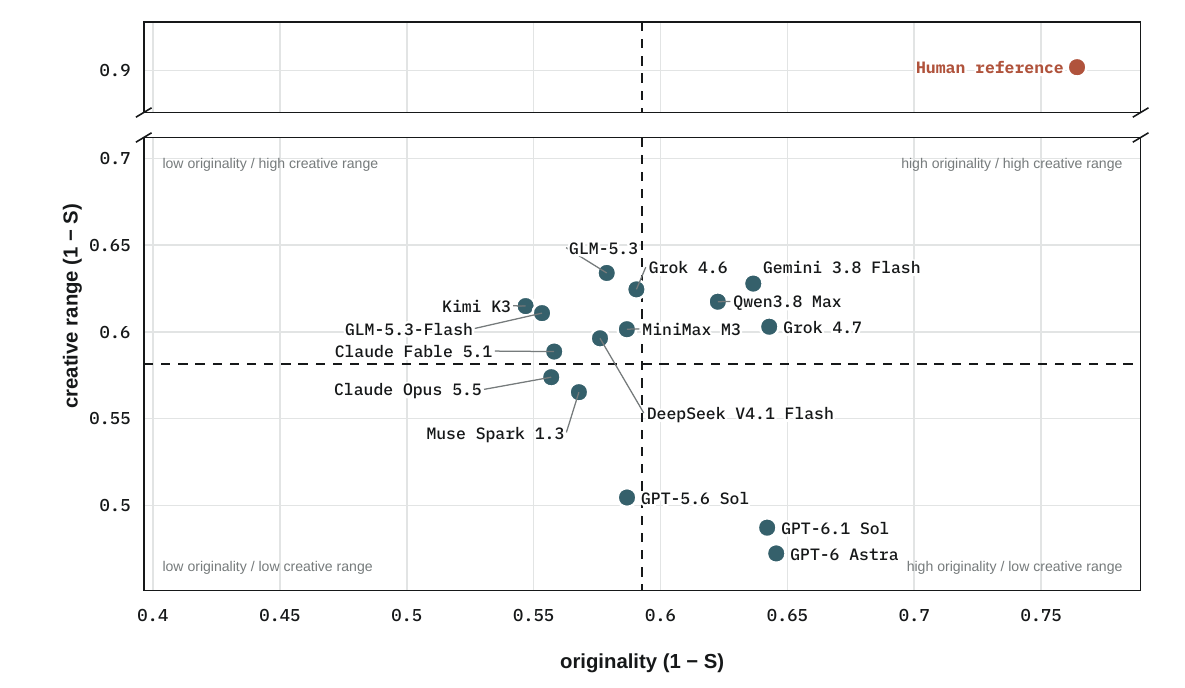}
\caption{\textbf{Creative range against originality.}
Each point is one model: its originality, the mean distinctiveness $1-S$ of its same-prompt pairs with other models
(Table~\ref{tab:diversity}; increasing to the right) against its creative range, the mean
distinctiveness of its two-prompt pairs (Table~\ref{tab:creativity}; increasing upwards). Both axes
are scaled so that one standard deviation of the fifteen models spans the same length, and the
dashed lines at the model means (0.593 and 0.581; SD 0.036 and 0.052) split the plane into four
comparable quadrants. The human reference is placed on the same scale but excluded from the mean and
SD; it lies $4.8$ SD to the right and $6.1$ SD above, and is drawn above the break in the vertical
axis. Grok 4.7, Gemini 3.8 Flash and Qwen3.8 Max fall in the upper-right quadrant among
models: above-average originality and creative range.}
\label{fig:creativity-repetition}
\end{figure}

\paragraph{Originality does not imply creative range.}
The two rankings are unrelated (Spearman $-0.13$, $p=0.66$; exploratory 95\% model-bootstrap
interval $[-0.65, 0.50]$, 20,000 resamples, seed 7; Figure~\ref{fig:creativity-repetition}).
GPT-6 Astra is the most original model in Table~\ref{tab:diversity} and last in
Table~\ref{tab:creativity}, differing from its competitors while repeating its own design on most
domain changes; GPT-6.1 Sol and GPT-5.6 Sol, second and third last, show the same pattern. Grok 4.7,
Gemini 3.8 Flash and Qwen3.8 Max are above the model mean on both. Originality and creative range
measure different properties; these rankings do not establish statistical independence.

\paragraph{Design outputs of today's frontier models are generally appropriate.}
As reported in Table~\ref{tab:appropriateness}, every model in our current test set meets more than 90\% of its briefs' criteria, and every model meets more than 96\% of the critical criteria. Claude Opus 5.5 scored the highest mean criteria satisfaction at 99.2\%, followed by GPT-6 Astra at 99.1\%, both slightly above the human reference at 98.0\%. The most common failures related to appropriateness are missing controls for a workflow the brief names (33\%), or sample data whose counts and totals do not match (26\%).

\begin{table}[htbp]
\centering
\small
\caption{Appropriateness. The mean pass rate is the percentage of a brief's criteria a design meets,
averaged over a model's designs; critical criteria met is the percentage of critical criteria met across all of a model's designs. Scores use every explicit criterion and the implied criteria of critical or major priority. Both are computed on all 160 prompts. Bold marks the highest model value in each column.}
\label{tab:appropriateness}
\begin{tabular}{@{}rlrr@{}}
\toprule
Rank & Model & Mean pass rate & Critical criteria met \\
\midrule
1 & Claude Opus 5.5 & \textbf{99.2\%} & \textbf{99.9\%} \\
2 & GPT-6 Astra & 99.1\% & 99.7\% \\
3 & GPT-6.1 Sol & 98.8\% & 99.6\% \\
4 & Claude Fable 5.1 & 98.5\% & 98.9\% \\
5 & GPT-5.6 Sol & 98.2\% & 99.3\% \\
6 & Grok 4.7 & 97.8\% & 98.5\% \\
7 & Muse Spark 1.3 & 97.8\% & 99.0\% \\
8 & Qwen3.8 Max & 97.6\% & 98.1\% \\
9 & Gemini 3.8 Flash & 97.5\% & 98.7\% \\
10 & Kimi K3 & 96.8\% & 97.7\% \\
11 & DeepSeek V4.1 Flash & 96.3\% & 98.0\% \\
12 & GLM-5.3-Flash & 96.1\% & 97.2\% \\
13 & GLM-5.3 & 95.6\% & 97.2\% \\
14 & Grok 4.6 & 95.5\% & 97.4\% \\
15 & MiniMax M3 & 93.4\% & 96.4\% \\
\midrule
--- & Human reference & 98.0\% & 98.6\% \\
\bottomrule
\end{tabular}
\end{table}

\begin{figure}[htbp]
\centering
\includegraphics[width=0.8\textwidth]{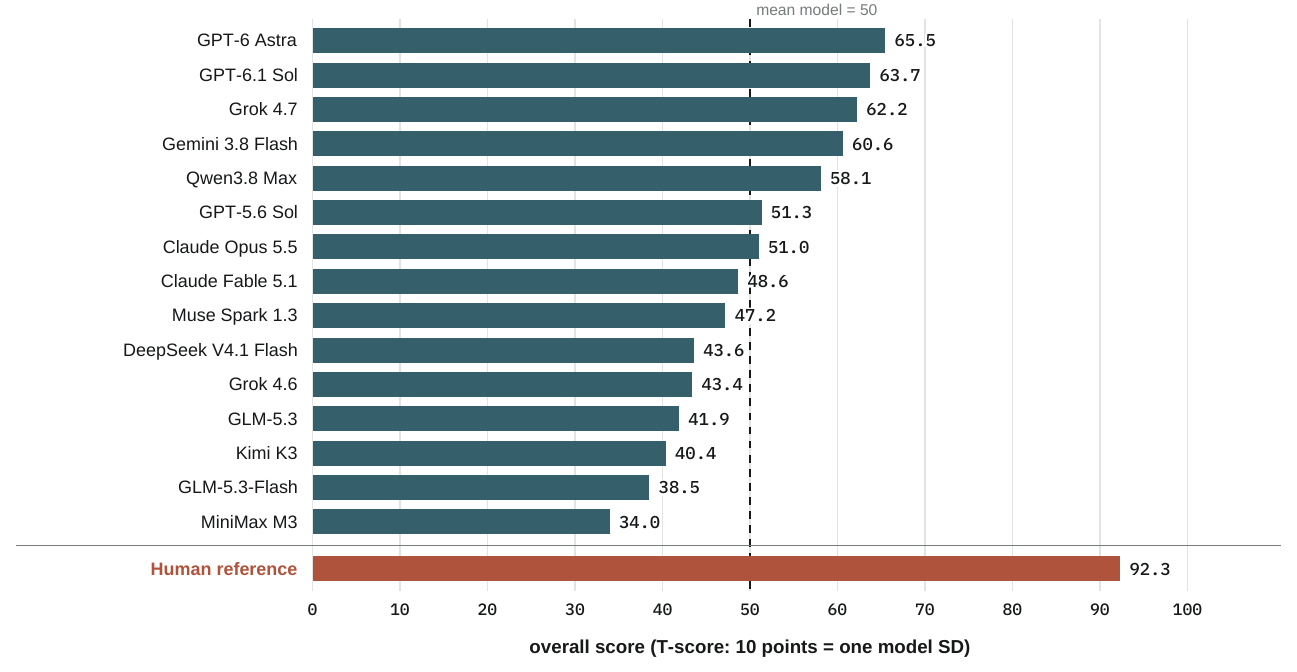}
\caption{\textbf{Overall score.} The mean of standardised diversity ($0.9\times$ originality
$+\,0.1\times$ creative range) and standardised appropriateness, standardised again and shown as a
T-score: 50 is the mean model and 10 points one model standard deviation. The human reference is placed on the same scale but
excluded from the mean and standard deviation. Appendix~\ref{app:overall} gives the details.}
\label{fig:overall}
\end{figure}

\paragraph{Overall score.}
Figure~\ref{fig:overall} ranks the models on one overall score. The score gives equal weight to
appropriateness and to diversity. Diversity combines originality and creative range with weights of
0.9 and 0.1; Appendix~\ref{app:overall} explains the weights. The score is a T-score: 50 is the
average model, and 10 points is one standard deviation. GPT-6 Astra (65.5) and GPT-6.1 Sol (63.7)
rank highest among models below the human reference score of 92.3.

\section{Limitations}

We do not control for temperature, sampling or other decoding parameters.
Hosted APIs are called with default temperature and sampling.

Repetition within one context is not measured: each design is generated in a fresh context, so we
do not generate a second design while the first is still visible to the model.

No diversity-seeking prompting was attempted: all models receive the same prompt without any specialised techniques to invoke creative outputs.

The human reference comprises designs created by multiple human designers, who adapted references from public inspiration libraries to the benchmark briefs. It supports comparisons between model outputs and human-designed solutions to the same briefs, as well as comparisons between human designs across paired briefs. However, it does not measure same-designer repetition or variation between different human designers responding to the same prompt.

\section{Conclusion}

Today’s frontier models generally produce UI designs that meet our appropriateness requirements. They nevertheless fall well short on diversity measures. Under the tested generation settings, designs from different models resemble one another more than they resemble the human reference, and every model exhibits less variation across paired briefs than the human reference does. Even the most original model is 0.12 below the human reference in originality, while the model with the widest creative range is 0.27 below it.

Originality and creative range capture different aspects of this limitation. GPT-6 Astra has the highest originality but the narrowest creative range: its designs differ from those of other models while remaining comparatively similar across changes in product domain and content. A distinctive model style can therefore coexist with limited variation across briefs.

Our work extends the study of LLM output diversity to UI design and examines repetition both across models and across related briefs. The human baseline shows that substantially greater design diversity is compatible with high appropriateness on these tasks. Design Creativity Bench provides a framework for measuring these differences and evaluating techniques that increase the diversity of LLM-generated designs while preserving their ability to meet the brief’s requirements. A live leaderboard with newer models is maintained at \url{https://kombai.com/benchmark/}.

\section*{Competing interests}

All authors are affiliated with Kombai Inc., which develops AI products for UI design and front-end development.

\clearpage
\appendix
\section{Supplementary methods and results}
\renewcommand{\thesubsection}{\thesection\arabic{subsection}}

\subsection{How much one prompt leaves open}
\label{app:entropy}

Section~\ref{sec:intro} claims that a UI prompt leaves a large space of good answers. If that is
true, two independent designers have little reason to arrive at the same screen. This appendix
illustrates the claim on a single prompt. The prompt is a separate illustrative example written for
this appendix, not one of the $\dbench$ prompts, and it is not representative of them: unlike the
benchmark prompts, it fixes navigation, palette, type scale and components. We enumerate a set of screens a competent designer could
defensibly build for it. The enumeration is illustrative, not exhaustive. We then ask how much
choice remains once everything the prompt fixes is set aside.

\paragraph{The prompt.}
The prompt asks for a \emph{Staff $\to$ Access} screen for Ridgewell Clinics, a fictional regional
outpatient provider. The screen lives inside a staff application on a desktop viewport. An
administrator uses it to give each person exactly one global role. The administrator also uses it
to inspect what a role is allowed to do. The content is fixed in advance and is deliberately
untidy: seven staff, five roles, twelve permissions and no per-person exceptions. Location is
supplied as workplace context and grants no access. The prompt also fixes the surrounding
application and its existing navigation, the desktop viewport, and the design system that supplies
the colours, the type scale and the components.

\paragraph{What the prompt leaves open.}
Five questions remain. How is the screen divided into panes? Which object organises it, the person
or the role? What does the administrator see before acting? How are the twelve permissions
revealed? How is an edit committed?

\begin{figure}[htbp]
\centering
\includegraphics[width=\textwidth]{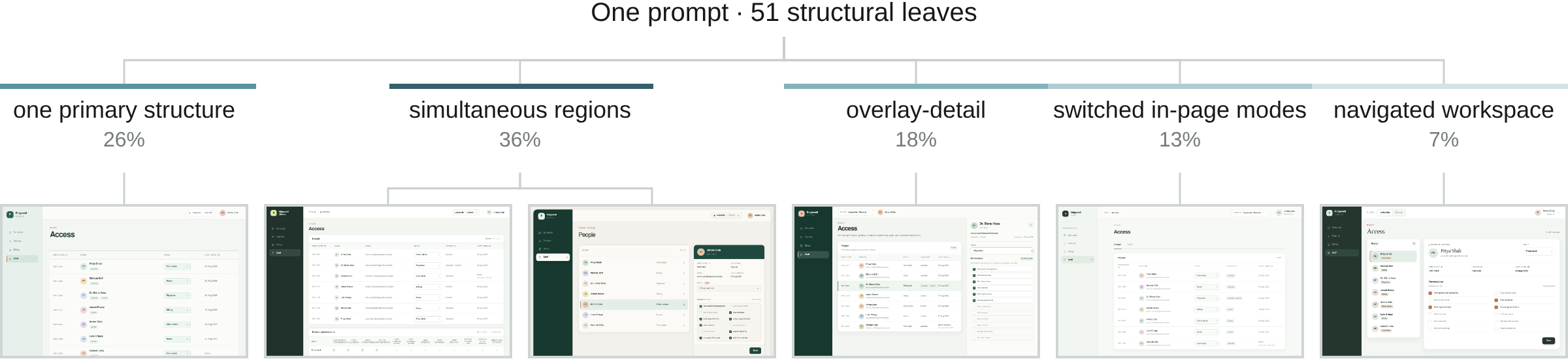}
\caption{\textbf{Part of the design space of one prompt.} The five top-level forks of the
enumeration, with the share of the probability mass under each. Six of the 51 leaves are shown as
rendered screens. Every branch shows its likeliest leaf. The largest branch also shows its second
likeliest, so that a fork inside a branch is visible as well. From the left: a people table with permissions
on a hover menu (3.9\%); a roster beside a roles-by-permissions map (5.6\%); a people list with a
role inspector (3.7\%); a roster that opens a person drawer (8.3\%); tabs that switch between the
people table and the role map (3.9\%); and a roster that opens a dedicated person page (4.3\%).
These six carry 29.7\% of the mass between them. The prompt fixes the navigation, the palette and
the components, so the variation to look for sits in the work area.}
\label{fig:design-tree}
\end{figure}

\paragraph{Counting the options.}
We enumerated 63 complete screens that answer those five questions defensibly. Other defensible
answers may exist. Some of them differ only in runtime state. A filtered table, an empty search, a failed row
and a save confirmation are one design seen at four moments. Folding those into their parent leaves
51 distinct structures. We call each one a leaf (Figure~\ref{fig:design-tree}). The enumeration is a partition, so every enumerated design
belongs to exactly one branch and none is counted twice. Each leaf also carries a probability. The
probability is the product of the decisions along its path. Each decision is weighted by how
conventional an LLM judges it to be, so familiar structures come out more likely than unusual ones.
These weights are assumptions. They are not estimated from choices made by real designers, and
every probability below depends on them.

\begin{figure}[htbp]
\centering
\includegraphics[width=\textwidth]{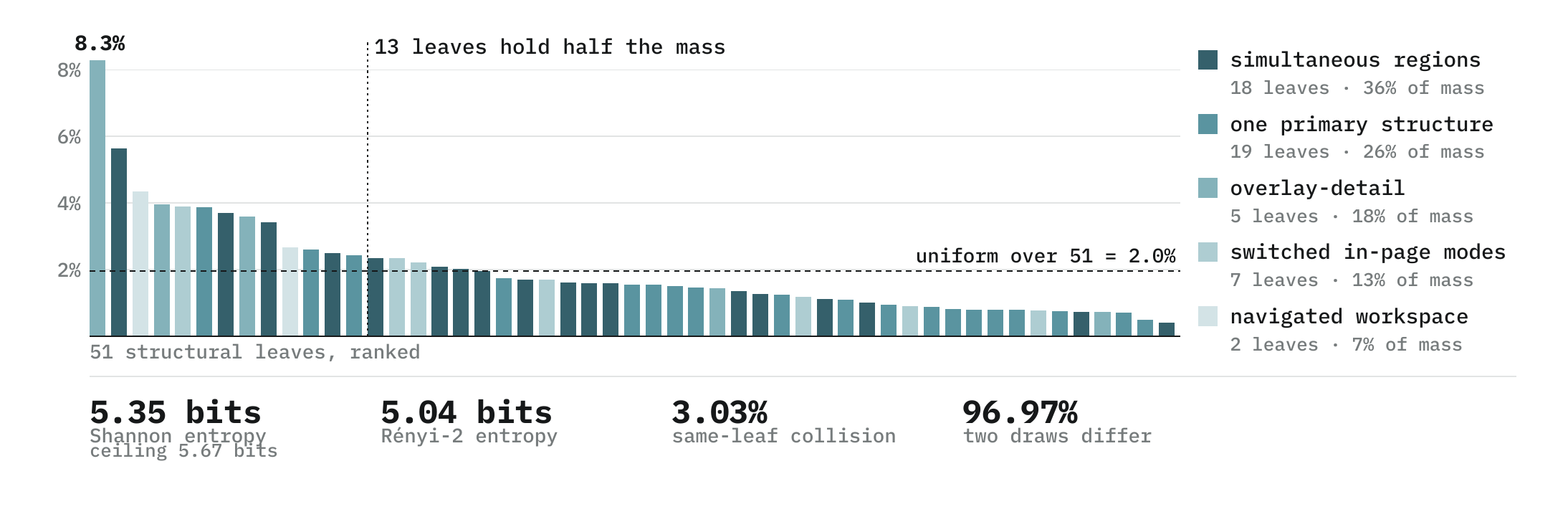}
\caption{\textbf{Residual design space of one prompt.} The 51 structural leaves of the
Staff~$\to$~Access prompt, ranked by probability and coloured by the top-level branch of the
enumeration. Each leaf's probability is the product of the decisions along its path, weighted so
that conventional structures come out more likely (an LLM determines the weights). The dashed line is what a uniform choice among
51 leaves would give. The distribution is close to flat. The most likely structure, a roster that
opens a drawer for one person, takes 8.3\% of the mass, and thirteen leaves are needed to reach
half of it. The collision probability $\sum_i p_i^2$ is 3.03\%, so two independent draws from
this assumed distribution differ in structure 96.97\% of the time. This follows from the weights;
it is not an observed rate for designers.}

\label{fig:design-entropy}
\end{figure}

\paragraph{How flat the distribution is.}
Under the assumed weights, the resulting distribution is close to flat (Figure~\ref{fig:design-entropy}). Its Shannon entropy
is 5.35 bits. A uniform choice among 51 leaves would give $\log_2 51 = 5.67$ bits. The most likely
structure takes only 8.3\% of the mass. The tenth most likely structure still takes 2.7\%.
Thirteen leaves are needed to cover half the mass.

\paragraph{What this means for a pair of designs.}
For a benchmark built on pairs, the quantity of interest is the collision probability
$\sum_i p_i^2$. It is the chance that two designers drawing independently from this distribution
land on the same structure. Here it is 3.03\%. Two independent competent designs to this prompt
therefore differ in structure 96.97\% of the time. The same fact stated as an entropy is a
R\'enyi-2 entropy of 5.04 bits. Over 500 independent pairs, the standard deviation of that 96.97\%
rate is 0.008.

\section{Evaluation dataset}
\label{app:prompts}

\paragraph{Tasks, prompts, and pairs.}
Figure~\ref{fig:corpus} shows a sample of three prompt pairs from $\dbench$ across design archetypes and verticals.
The pairs are constructed such that they share the same design goal and UI archetype.
Each of the 80 tasks is written for two product domains, giving
$80\times2=160$ prompts. The two prompts of a task form its differentiated pair,
giving 80 pairs per model.
No decisions related to visual styles or information architecture are mentioned in any of the prompts, and they are left entirely up to the model
to decide (along with the majority of the content). Brand name, domain taxonomy and some key content serve as differentiating attributes between the prompts of a task.
As illustrated in Appendix~\ref{app:entropy}, our prompt construction leaves many design decisions up to the model.

\paragraph{Generation.}
Each prompt is passed to the model through the same wrapper, which requests a complete standalone
HTML document submitted through a single tool. The output length distribution of the HTML documents is shown
in Figure~\ref{fig:corpus}, bottom right panel. Models are asked to populate realistic mock data in
tables, lists, charts, calendars, and detail panels, present at first paint without a data fetch or
user interaction. External assets, frameworks, and files are disallowed unless explicitly permitted
by the task. If the HTML raises an error when rendered in a browser, the trace is returned for at
most three repair attempts; sixteen outputs errored on the first attempt and all sixteen cleared on
the first repair.

We use native hosted APIs with default temperature and sampling settings and the reasoning
efforts in Table~\ref{tab:models}. 

\subsection{Segment extraction and image capture}
\label{app:segments}

Two things about a page are compared separately. The first is its information architecture (IA):
how the page is organised and composed. The second is its colour system. IA is compared through
segments, which this appendix defines. Colour is compared as a property of the whole page.

\paragraph{What a segment is.}
A segment is a container in the rendered page. It holds at least two boxed children and covers at
least 6,000 pixels of area. A boxed child is a child element whose rendered width and height both
exceed one pixel. Segments are the compositions a reader would name: a navigation bar, a table, a
content panel. One segment can sit inside another. The segments of a page are therefore overlapping
regions rather than a partition of the screenshot. Text and icons are not segments on their own.
They are represented through the composition that contains them.

\paragraph{How segments are found.}
Detection is deterministic and calls no model. Every DOM element is given an identifier. Each
element's box and visibility are then measured in the browser, and the elements that meet the
definition above are retained. Hidden elements are excluded. Media elements are excluded as
containers: \texttt{img}, \texttt{video}, \texttt{svg}, \texttt{canvas}, \texttt{picture},
\texttt{iframe} and \texttt{object}. 

\paragraph{What the segments feed.}
A segment's area and its position relative to the initial viewport set how much the page score
weights it. A page can hold the same composition more than once. Such a repeat is
removed when it finds no counterpart on the other page and its own page already has a matched
copy of it. The weights are recomputed over what remains. The colour system is read from the same rendered page. Colour
invariance is learned by the projection during training, so the page itself is left unmodified.
Appendix~\ref{app:scoring} defines the encoder calibration, the size blend, the importance
weights, the matching between two pages' segments, and the score built from them.

\subsection{Projection dataset, training, and metrics}
\label{app:projection}

Some edits to a page change how it looks without changing the design. A new colour theme, a new
copy string and a wider viewport are examples. We call these nuisance edits. Other edits change the
design itself. Deleting a child, changing the layout axis and reparenting an element are examples.
We call these structural edits. A similarity score is useful here when it ignores the first kind
and reacts to the second. The two encoders here are gemini-embedding-2 and UIClip ViT-B/32.
Raw cosine similarity in either of their embedding spaces separates the two
kinds of edit barely above chance (Table~\ref{tab:projection}). We therefore fit one
compact linear map per encoder on a dataset built from these two kinds of edit.

\paragraph{The dataset.}
The projection dataset $\dproj$ is different from the benchmark $\dbench$. It is obtained from pages generated from a mix of
GPT-5.6 Luna, Qwen 3.7 Flash, Gemini 3 Flash Preview and MiniMax M3. Product domain, objective, audience, page shell, interaction, state, density, content mix and
section count change among these pages. Every page is
written with CSS theme variables and ARIA roles, which lets an edit be applied in the browser.

\paragraph{Training pairs.}
The builder takes up to sixteen containers from each page as anchors. An anchor is captured twice,
once before an edit and once after it. Those two images are one training pair. The pair is labelled
by the family of the edit. Nuisance families are viewport change, copy substitution, theme change,
spacing, responsive reflow, type scale, wrapper depth, icon or graphic change, and zoom. The UIClip
variant adds chrome, tone and letterform changes. Structural families delete or clone children,
keep only the first child, change the layout axis, remove layout rules, reorder children, flatten
the hierarchy, or reparent elements. A structural edit whose screenshot does not visibly change is
rejected. Every record stores the source page, the anchor, the edit family, both images, the
geometry and a validity flag.

The Gemini manifest holds 86,238 valid pairs: 58,805 nuisance and 27,433 structural. The UIClip
manifest holds 104,428: 77,000 nuisance and 27,428 structural.

\paragraph{The map.}
Pages are shuffled under a fixed seed and split 80/20, which keeps every anchor and every variant
of a page on one side. This gives 304 training pages, $\dptrain$, and 77 validation pages,
$\dpval$. The input for encoder $e$ is the crop's embedding $x_e$ with standardised log-width and
log-height appended. The map is
\begin{equation}
 z_e(x_e)=W_e\bigl([x_e;\widehat{\log w};\widehat{\log h}]-\mu_e\bigr),
 \label{eq:projection}
\end{equation}
where $\mu_e$ is the centroid of $\dptrain$ and $W_e$ has 64 output dimensions.

\paragraph{The loss.}
Each comparison is a triplet drawn from one anchor: the crop itself, the same crop after a nuisance
edit, and the same crop after a structural edit. The map is asked to hold the nuisance variant
closer than the structural one by a fixed margin. Learning a metric from relative comparisons in this way is standard
\citep{schroff2015facenet}. Write $c$ for cosine similarity between projected vectors, $p$ for a
nuisance variant, $n$ for a structural variant, $(u,v)$ for two crops from different pages, and
$[x]_+=\max(x,0)$. The loss contains
\begin{equation}
\begin{split}
 \mathcal{L}_{\rm rank}&=\mathrm{E}_{(a,p,n)}[0.15+c(a,n)-c(a,p)]_+,\\
 \mathcal{L}_{\rm inv}&=\mathrm{E}_{(a,p)}[0.90-c(a,p)]_+,\\
 \mathcal{L}_{\rm cal}&=\mathrm{E}_{(u,v)}c(u,v)^2,\qquad
 \mathcal{L}_{\rm orth}=0.05\lVert WW^\top-I\rVert_F^2.
\end{split}
\label{eq:projection-loss}
\end{equation}
The first term carries the margin. The second asks a nuisance variant to sit at similarity 0.90 or
above. The third pulls crops from different pages towards zero. The fourth keeps the rows of $W$
close to orthonormal. Pairs that are already near-duplicates in the original embedding space are
excluded from the third term. Training runs for 60 epochs at learning rate 0.05, in batches of 256
nuisance pairs together with the structural comparisons that share their anchors. The checkpoint
with the best AUC on $\dpval$ is kept.

\paragraph{Diagnostics.}
Pair accuracy is the share of triplets in which the nuisance variant scores above the structural
variant from the same anchor. AUC compares the nuisance and structural similarity distributions
over all anchors. The three medians describe the scale the score works on. Table~\ref{tab:projection}
reports all of these on $\dpval$. $\dpval$ also chose the checkpoint, so they are model-selection
diagnostics rather than a held-out test. 

\begin{table}[htbp]
\centering\small
\caption{Projection diagnostics on $\dpval$, the split that also selected the checkpoint. Nuis.,
str.\ and unrel.\ are the median similarities of nuisance pairs, structural pairs, and pairs of
crops from different pages.}
\label{tab:projection}
\begin{tabular}{@{}lrrrrr@{}}
\toprule
Space & Pair acc. & AUC & Nuis. & Str. & Unrel. \\
\midrule
Gemini & .645 & .640 & .980 & .966 & .074 \\
Gemini, projected & .732 & .722 & .945 & .879 & $-.016$ \\
UIClip & .661 & .649 & .996 & .990 & .008 \\
UIClip, projected & .730 & .724 & .954 & .862 & $-.019$ \\
\bottomrule
\end{tabular}
\end{table}

\subsection{Similarity calibration and score implementation}
\label{app:scoring}

This appendix explains the similarity score for two UI segments. Appendix~\ref{app:segments} defines the
segments and their images. Appendix~\ref{app:projection} defines the two projected spaces, the
nuisance and structural edits they are fitted on, and the validation split $\dpval$ that the
calibrations here are read from.

\paragraph{Putting the two encoders on one scale.}
Each segment image supplies one projected vector per encoder. For segments $i$ and $j$ from the two
pages we take the cosine similarities $c^G_{ij}$ and $c^U_{ij}$ in the projected Gemini and UIClip
spaces. The two spaces do not share a scale. Let $u_e$ and $n_e$ be the median similarities of
unrelated pairs and of nuisance pairs for encoder $e$ on $\dpval$. UIClip is mapped onto Gemini's
scale by
\begin{equation}
 \widetilde{c}^U_{ij}=u_G+
 \frac{n_G-u_G}{n_U-u_U}\bigl(c^U_{ij}-u_U\bigr).
 \label{eq:encoder-calibration}
\end{equation}
Unrelated pairs now sit at the same baseline in both spaces, and nuisance pairs sit at the same
high reference. A single matching threshold ($\tau$, below) therefore means the same thing in both.

\paragraph{Creating the hybrid embedding.}
UIClip's weight increases with segment area, while Gemini retains the majority of the weight for regions smaller than one reference viewport. With segment
areas $a_i,a_j$, reference viewport area $V=1280\times720$, and the logistic function
$\sigma(t)=1/(1+e^{-t})$,
\begin{equation}
 \alpha_{ij}=\sigma\!\left(\frac{\log(\min(a_i,a_j)/V)}{0.5}\right),\qquad
 s_{ij}=(1-\alpha_{ij})c^G_{ij}+\alpha_{ij}\widetilde{c}^U_{ij}.
 \label{eq:similarity}
\end{equation}
Writing $r=\min(a_i,a_j)/V$, this gives $\alpha=r^2/(1+r^2)$: UIClip receives 0.99\% of the weight at $r=0.1$, 20\% at $r=0.5$, and 50\% at $r=1$. UIClip dominates only when both segments exceed one reference viewport in area.

\paragraph{Importance and the page budget.}
Models differ substantially in how much HTML and content they produce for one prompt
(Figure~\ref{fig:corpus}). A verbose model fills a page with extra sections and repeated records. A
terse model designs the same task with fewer elements. Counting matched segments would confuse
repetition with this difference in output size. We measure the \emph{fraction of important design
mass that repeats} instead. A page carries a capped quantity of importance spread over its elements.
Adding an element divides that quantity differently rather than adding to it. Segment-wise scoring
can then compare how two pages allocate importance even when they hold very different numbers of
segments.

A segment's intrinsic importance follows from the space it occupies and from whether the reader
meets it on arrival. For segment area $a_i$ and reference viewport area $V=1280\times720$, the viewport at which pages are
rendered and measured, it is
\begin{equation}
 r_i=\min(a_i/V,1)\,p_i,\qquad
 p_i=\begin{cases}
 1 & \text{in the initial viewport},\\
 0.6 & \text{below the fold},\\
 0.1 & \text{overlay or conditional content}.
 \end{cases}
 \label{eq:importance}
\end{equation}
Content outside the initial view is capped at half of each page's total. Total page importance is
capped at $\mathcal{B}=5.4$. The same budget then applies to every
comparison. Writing $R_{\rm in}$ and $R_{\rm out}$ for the intrinsic importance inside and outside
the initial view, the final weights are
\begin{equation}
 \kappa=\min(1,R_{\rm in}/R_{\rm out}),\quad
 r'_i=\begin{cases}r_i& i\text{ inside},\\ \kappa r_i&i\text{ outside},\end{cases}
 \qquad w_i=r'_i\min\!\left(1,\frac{\mathcal{B}}{\sum_k r'_k}\right),
 \label{eq:budget}
\end{equation}
with $\kappa=1$ when $R_{\rm out}=0$. Each page's weights are computed on its own. Adding sections
cannot raise a page's importance without bound.

\paragraph{Matching the two segment sets.}
Let $A$ and $B$ denote the two segment sets. The Hungarian algorithm
\citep{kuhn1955hungarian,munkres1957algorithms} gives a partial one-to-one matching $M^*$, with a
null assignment available for every segment:
\begin{equation}
 M^*=\mathop{\arg\min}_{M\ \text{one-to-one}}
 \left[\sum_{(i,j)\in M}(1-s_{ij})+
 \frac{1-\tau}{2}\bigl(|A|+|B|-2|M|\bigr)\right],\qquad \tau=0.45.
 \label{eq:matching}
\end{equation}
A match is preferred to leaving both segments unmatched only when its similarity clears $\tau$.
We selected $\tau$ on $\dpval$ from 6,094 nuisance pairs and 4,000 unrelated pairs.

The assignment is solved with the rectangular variant of \citet{crouse2016implementing}. The cost
matrix holds the segment costs $1-s_{ij}$, one null slot per segment at cost $(1-\tau)/2$, and
null--null assignments at zero cost. Unrelated segments are therefore never forced to match.

After matching, an unmatched segment that near-duplicates a matched segment on its own page is
removed at similarity 0.90. Both pages' weights are then recomputed over the segments that remain,
giving the weights $w_i$ used below. A second copy of a pattern the page already shows is
therefore not counted as an unmatched new composition.

\paragraph{The IA score.}
Writing $s_{ij}$ for the blended similarity of segments $i$ and $j$, and $w_i$ for those weights,
the IA score is
\begin{equation}
 S_{\rm IA}(A,B)=
 \frac{\sum_{(i,j)\in M^*}\min(w_i^A,w_j^B)\,s_{ij}}
 {\max\!\left(\sum_{i\in A}w_i^A,\ \sum_{j\in B}w_j^B\right)}.
 \label{eq:ia}
\end{equation}

Dividing by the larger of the two page totals puts the result on a common scale. Writing
$T_A=\sum_i w_i^A$ and $T_B=\sum_j w_j^B$, each non-empty page has a unit-mass allocation
$\pi_i^A=w_i^A/T_A$ and $\pi_j^B=w_j^B/T_B$, so $\sum_i\pi_i^A=\sum_j\pi_j^B=1$. The allocation
says where a page's importance lies, independently of how many segments it holds. Let
$\rho_A=T_A/\max(T_A,T_B)$ and $\rho_B=T_B/\max(T_A,T_B)$. The implemented score can be written
equivalently as
\begin{equation}
 S_{\rm IA}=\sum_{(i,j)\in M^*}
 \min(\rho_A\pi_i^A,\rho_B\pi_j^B)s_{ij}.
 \label{eq:unit-mass}
\end{equation}
This allows us to score pairs of pages with completely different sizes and numbers of segments.

\subsection{Colour system and the final score}
\label{app:colour}

Colour is a property of the page as a whole. A shared palette should count once, however many
segments use it. Humans also weigh colour differently from composition. Colour is therefore scored
separately and added to $S_{\rm IA}$ at a fixed weight.

\paragraph{Extraction.}
An LLM pass reads each page's source CSS and returns its palette values in a structured schema of
six roles. The extractor is instructed to copy colours from CSS custom properties or literal
declarations, to omit roles the page does not express, and to exclude content-specific tints such
as per-person avatar colours. The role set and the role weights in Table~\ref{tab:colour-weights}
are selected by hand. The LLM supplies the page-specific values alone.

\begin{table}[htbp]
\centering\small
\caption{Hand-selected colour-system keys and relative weights.}
\label{tab:colour-weights}
\begin{tabular}{@{}llr@{}}
\toprule
Key & Page role & Weight \\
\midrule
\texttt{bg} & Page background & 1.0 \\
\texttt{surface} & Panel or card background & 1.0 \\
\texttt{ink} & Primary text & 1.0 \\
\texttt{ink\_muted} & Secondary or muted text & 0.5 \\
\texttt{accent} & Primary interactive accent & 1.0 \\
\texttt{border\_color} & Hairline or divider colour & 0.5 \\
\bottomrule
\end{tabular}
\end{table}

The four primary roles carry full weight. Muted text and borders carry half, because they are
usually derived from the primary palette. The six weights sum to five. A role the page omits falls
back to a fixed colour, so all six keys are always present and the ceiling stays at five.

\paragraph{Comparing two palettes.}
Each role is compared deterministically. HSL hue $H$ and lightness $L$ give a chroma vector
$q=C(\cos 2\pi H,\sin 2\pi H)$, where $C=S(1-|2L-1|)$ and $S$ is HSL saturation. Hue disagreement
is the distance between the two vectors divided by their summed chroma. It is attenuated once that
sum falls below 0.3. Role similarity is one minus the larger of hue disagreement and absolute
lightness difference.

\paragraph{The final score.}
With role similarities $c_k$ and role weights $v_k$,
\begin{equation}
 S_{\rm colour}=\frac{\sum_k v_k c_k}{\sum_k v_k},\qquad
 S(A,B)=0.9\,S_{\rm IA}+0.1\,S_{\rm colour}.
 \label{eq:final}
\end{equation}

\subsection{Human validation and scoring alternatives}
\label{app:human}

A scoring method has to meet two requirements. It has to agree with people about which designs
repeat, and it has to move consistently when a design is deliberately changed. This appendix covers
the first requirement, using the human judgements in $\dhuman$. Appendix~\ref{app:ladders} covers
the second, using the controlled perturbations in $\dladder$.

\paragraph{The study.}
$\dhuman$ holds 100 cases. A case shows a reviewer two pairs of designs. The reviewer picks the
pair whose two designs are more similar to each other. A reviewer who cannot separate them may
call a tie. Three reviewers see all 100 cases. Reviewers are asked to rank one pair against
another rather than to rate a pair on a scale, because a relative judgement is the easier one to
make consistently. These labels are therefore relative. 

\paragraph{Pooling the labels.}
The three reviewers together supply one pooled label per case. Where their non-tied choices agree,
that choice stands. Where those choices split evenly, the pooled label is a tie. A tie from one
reviewer does not override a choice from another. This leaves 80 cases with a non-tied pooled
label, and those 80 are what every scoring method is measured on.

The reviewers are also compared with each other. For each of the three reviewer pairs we take the
cases where both members made a non-tied choice, and measure how often they picked the same pair.
Averaged over the three reviewer pairs (82.1\%, 79.4\% and 93.4\%, on 56, 63 and 76 cases), agreement is 85.0\%. 

\paragraph{The judge protocol.}
The judge panel is Gemini 3.8 Flash, GPT-5.6 Luna, MiniMax M3, Qwen3.8 27B, Qwen3.8 Flash and
Grok 4.6. A judge is given the same case and picks a pair directly. A panel prediction needs at
least four non-tied votes, and split majorities are dropped. A single judge is scored on the cases
it answered without a tie. Judge denominators are therefore smaller than the 80 cases the embedding
methods use. A judge ranks a pair rather than scoring it, so no judge has a continuous score and no
judge has an AUC. 

\begin{figure}[t]
\centering
\includegraphics[width=\textwidth]{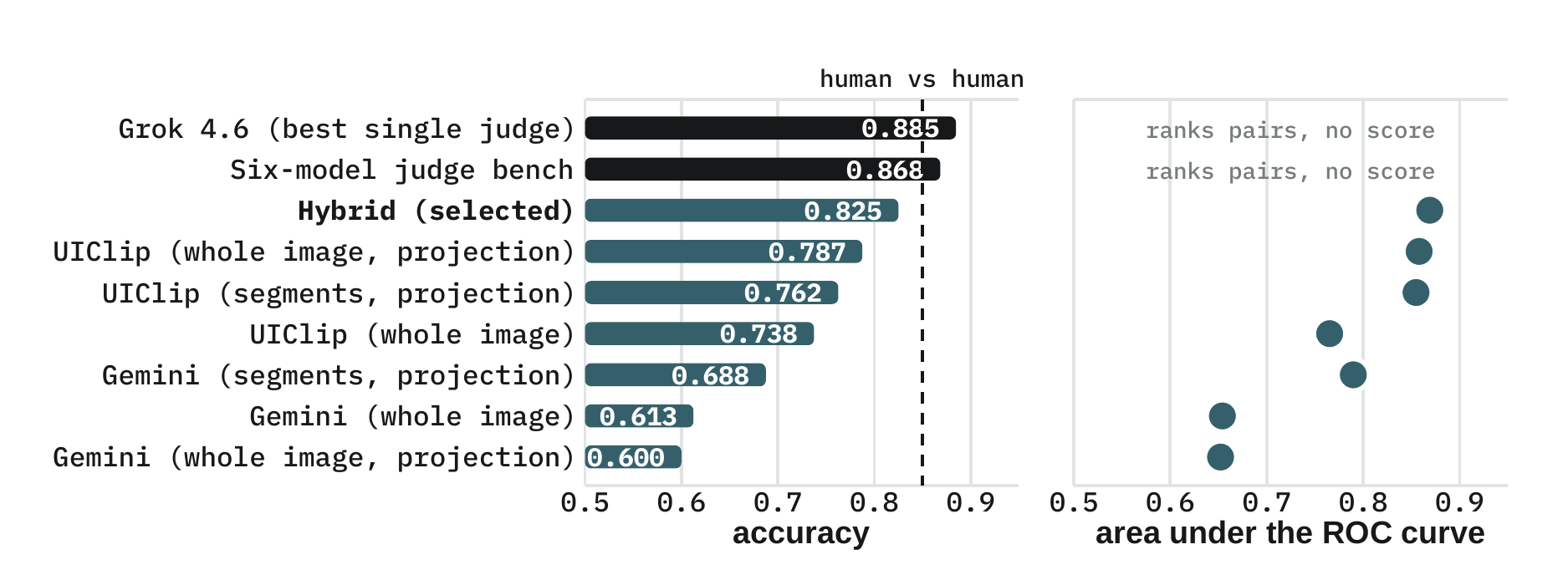}
\caption{\textbf{Does the score agree with people about design repetition?}
On $\dhuman$, reviewers choose which of two design pairs is more similar. Left: agreement with the
pooled human choice; right: ROC AUC from the numerical score difference, which the judges do not
have because they rank a pair rather than scoring it. Embedding methods use the 80 cases with a
non-tied pooled label; judge accuracies use their answered cases, excluding split majority votes.
The dashed line averages the three reviewer pairs, each measured on the comparisons where both of
its members made a non-tied choice (195 in total).
The hybrid segment metric has the highest embedding accuracy; both LLM judges have higher
agreement. Robustness to controlled changes is tested separately in Figure~\ref{fig:ladders}.}
\label{fig:human-agreement}
\end{figure}

\paragraph{Results.}
Figure~\ref{fig:human-agreement} and Table~\ref{tab:scorers} compare every method on these labels:
whole-page and segment embeddings, raw and projected embeddings, and the judges. For a numerical
scorer the predicted choice is the pair with the larger score, and the difference between the two
scores also gives an ROC AUC \citep{hanley1982meaning}.

\begin{table}[htbp]
\centering\small
\caption{Human agreement of every participant in Figure~\ref{fig:human-agreement}, on the 80
cases with a non-tied pooled label. A row names its encoder, how it aggregates a page, and whether
the embeddings pass through the fitted projection. Judges rank a pair rather than scoring it, so
they have no AUC. Their denominators are smaller because they omit their own ties and split
majorities.}
\label{tab:scorers}
\begin{tabular}{@{}lrrr@{}}
\toprule
Participant & Accuracy & AUC & Cases \\
\midrule
Grok 4.6 (best single judge) & .885 & --- & 78 \\
Six-model judge bench & .868 & --- & 76 \\
Hybrid (segments, projection), selected & .825 & .869 & 80 \\
UIClip (whole image, projection) & .787 & .858 & 80 \\
UIClip (segments, projection) & .762 & .855 & 80 \\
UIClip (whole image) & .738 & .765 & 80 \\
Gemini (segments, projection) & .688 & .790 & 80 \\
Gemini (whole image) & .613 & .654 & 80 \\
Gemini (whole image, projection) & .600 & .652 & 80 \\
\bottomrule
\end{tabular}
\end{table}

\subsection{Controlled ladders and robustness ablations}
\label{app:ladders}

\paragraph{Robustness to convergence and perturbation.}
In $\dladder$, a \emph{ladder} is a sequence of three cumulative edits, L1--L3, with a known
direction of change. This is a metamorphic test \citep{chen1998metamorphic,segura2016survey}: the
edits define which way similarity should move.
In a convergence study, a seed design is progressively edited towards a fixed reference, so
similarity to that reference should rise at each level. In a perturbation study, a design is
progressively edited away from its original, so similarity to the original should fall. These
experiments test sensitivity to design changes without requiring human similarity labels.

\begin{figure}[t]
\centering
\includegraphics[width=\textwidth]{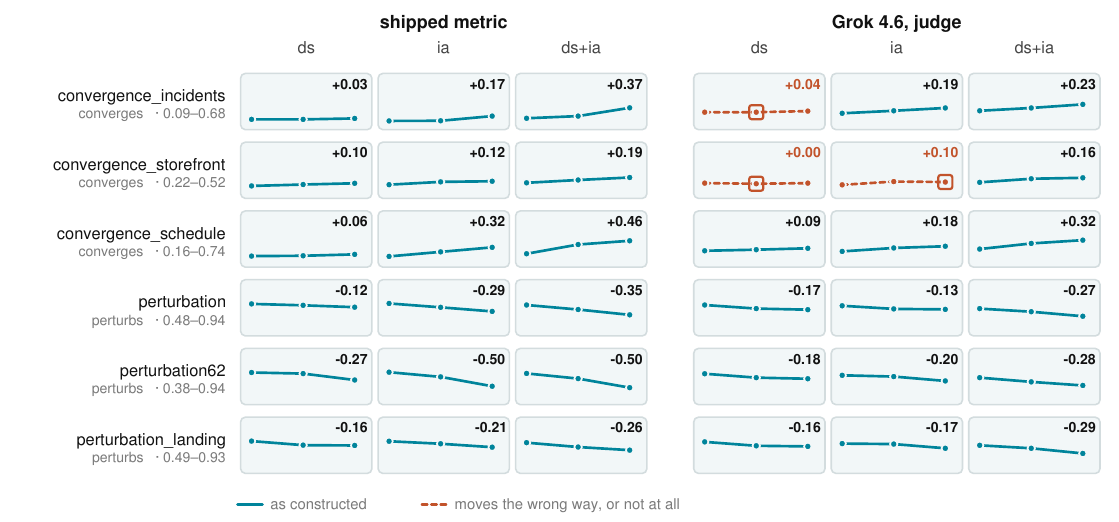}
\caption{\textbf{Does similarity move when the design changes?}
Each line in $\dladder$ follows three cumulative edit levels, L1--L3. Convergence edits move a seed towards a
fixed reference (scores should rise); perturbations move a design away from its original (scores
should fall). DS changes visual styling, IA changes content organisation, and DS+IA changes both.
Left: the selected metric; right: the strongest judge on human agreement, Grok 4.6. Each study
uses a shared vertical scale across its row; numbers give the score change from L1 to L3.
Dashed orange lines and boxes mark a plateau or reversal. The metric follows all 18 displayed
ladders; the judge follows 15.}
\label{fig:ladders}
\end{figure}

The labels in Figure~\ref{fig:ladders} specify what changes. A \emph{DS perturbation} changes the
visual styles---palette, typography, spacing, borders, shadows, and corner treatment---while
preserving information architecture. An \emph{IA perturbation} changes navigation, grouping,
layout, or the relationship between content regions while preserving the visual styles.
\emph{DS+IA} applies both sets of edits. The same labels in convergence studies indicate which
facets are moved towards the reference. L2 includes the L1 edits, and L3 includes both earlier
levels; the levels are ordered interventions.

Across three convergence and three perturbation studies in $\dladder$, the selected metric moves strictly in
the expected direction on all 18 DS, IA, and DS+IA ladders. Grok 4.6 does so on 15 of 18,
with plateaus or reversals in three convergence ladders. Figure~\ref{fig:ladders} shows these
controlled changes. Together, the two validation experiments motivate the segment metric:
it agrees reasonably with human judgements and responds consistently to the controlled changes
in the facets used by the benchmark.

The three convergence studies in $\dladder$ are incident management, a storefront, and scheduling.
Figure~\ref{fig:ladders} labels the three perturbation studies separately. Edits are specified
concretely for each seed: which palette values or style rules change and which regions move.
L2 applies the first two edit sets and L3 all three.

The available complete ladders total 18: nine convergence and nine perturbation ladders. A ladder
passes only if both adjacent score changes have the required strict sign; a plateau fails. The
scorer passes all nine convergence ladders and all nine perturbation ladders. Grok's three
failures are the incident DS, storefront DS, and storefront IA convergence ladders.

\begin{table}[htbp]
\centering\small
\caption{Robustness ablations on $\dladder$: complete DS, IA, and DS+IA ladders with strictly
monotone scores.}
\label{tab:robustness}
\begin{tabular}{@{}lr@{}}
\toprule
Configuration & Passing ladders \\
\midrule
Selected hybrid + colour & 18/18 \\
Hybrid + colour, geometric decay & 17/18 \\
Hybrid + thirteen theme fields, decay & 16/18 \\
Graded-colour projections & 16/18 \\
Gemini projected segments & 16/18 \\
UIClip projected segments & 17/18 \\
\bottomrule
\end{tabular}
\end{table}

\subsection{The $\dcalib$ agreement set}
\label{app:threshold}

\paragraph{How the hundred pairs were drawn.}
The pairs of $\dcalib$ were drawn from the pool of generated designs and shown blind. The reviewers saw no model
identity, no score and no comparison type, and the left--right order was randomised. The question
was whether the two designs would be accepted as two distinct designs, or whether one reads as a
rehash of the other. The reviewers judged composition and discounted the content domain.

\paragraph{Agreement.}
The similarity score separates the two verdicts with an AUC of 0.85. No threshold is fitted on
$\dcalib$: the benchmark reports the score itself, and $\dcalib$ serves only as a second check of
human agreement. Nothing in the scorer was selected or fitted on $\dcalib$ (the scoring configuration
and the structure/colour weight were chosen on $\dhuman$ and the ladders), so the AUC is an
out-of-sample check.

\subsection{The overall score}
\label{app:overall}

\paragraph{Diversity.}
The overall score first merges the two diversity measures into one. Diversity is
$0.9\times$ originality $+\,0.1\times$ creative range, on the raw $1-S$ scale.

The weights come from a simple model of how designs are used in practice. A person asks an AI agent
for a design. The person wants that design to differ from a typical AI-generated design already in
use. A typical design is a random draw from all AI-generated designs, so it comes from some model in
proportion to that model's usage. With probability $p$ the draw comes from the same model as the
person's agent, and with probability $1-p$ it comes from another model.

The two measures cover these two cases. If the typical design comes from the same model, creative
range measures how far apart the two designs are, because one model has designed against two different
briefs. If it comes from another model, originality measures it. The expected distinctiveness from a
typical design is therefore $p$ times creative range plus $1-p$ times originality. Creative range
takes the weight $p$, and originality takes the weight $1-p$.

If model $i$ produces a share $s_i$ of all AI-generated designs, the person's agent is model $i$ with
probability $s_i$, and the typical design is too. So $p=\sum_i s_i^2$. This sum is never larger than
the largest single share, $\max_i s_i$. So $p$ is small unless one model produces most of the
designs.

Shares of UI generation are not published. As a proxy we use the public token rankings of OpenRouter
for its twenty most-used models, retrieved on 30 September 2026.
There, the most-used model holds 19\% of the tokens, and $\sum_i s_i^2=0.10$. These shares are
computed within the top twenty only, so they overstate the true shares. OpenRouter also does not see
tools that call a provider directly, such as Claude Code and Codex. We set $p=0.1$.

The mix uses the raw scores, not standardised scores. Creative range varies more across models than
originality does (standard deviation 0.051 against 0.035). Standardising each measure first would
change how much each one moves the result, and the weights would no longer be shares of an expected
distinctiveness.

\paragraph{Combining with appropriateness.}
Diversity and appropriateness have different units, so each is standardised. We subtract the mean
of the fifteen models and divide by their standard deviation. We take the mean of the two standardised values, so each measure
carries half of the weight. The mean of two standardised values does not itself have unit standard
deviation: its variance is $(1+r)/2$, where $r$ is the correlation between the two. We therefore
standardise the mean once more over the models, and report the result as a T-score, $50+10z$. The
overall scores of the models then have mean 50 and standard deviation 10. Both steps are linear, so
they keep the ranking unchanged, and every model score stays positive. The human reference is
placed on the same scale but does not enter the mean or the standard deviation.

\paragraph{Sensitivity.}
The ranking is stable under nearby weights. With a creative-range weight of 0 or of 0.15 instead of
0.1, the rank correlation with the reported ranking is at least 0.99, and no model moves more than
two places. With a weight of 0.25 the correlation is 0.96: Grok 4.7 and Gemini 3.8 Flash move ahead
of GPT-6 Astra, and no model moves more than two places.

\subsection{Extended results with a frozen panel}
\label{app:extended}

The fifteen models of the paper form a frozen panel. A model added later is placed on the panel's
scale; it does not change any panel score. This appendix places eighteen further models on that
scale.

\paragraph{Frozen scoring.}
Three rules keep the panel's scores fixed.
\begin{itemize}
\item \textbf{Originality against the panel only.} A new model's originality is the mean
distinctiveness $1-S$ of its designs against the panel models' designs for the same prompt. New
models are not compared with each other, and panel models are not compared with new models. No new
model shares a lab with a panel model, so each new model has 15 partners and 2,400 pairs.
\item \textbf{Normalisation constants from the panel.} Diversity and appropriateness are
standardised with the panel's mean and standard deviation (diversity 0.5917 and 0.0307;
appropriateness 97.21\% and 1.59 points). The mean of the two standardised values is standardised
again with the panel's constants (mean 0 and standard deviation 0.773), and reported as $50+10z$.
\item \textbf{The same pipeline.} The briefs, the generation harness, the segmenter, the embedding
model, the judge and the per-brief criteria are the ones the panel was scored with.
\end{itemize}

\begin{table}[p]
\centering\small
\caption{Extended results on the frozen panel. The fifteen panel models keep the scores reported in
the paper; the eighteen new models (italic, with their reasoning effort) are placed on the panel's
scale. Originality of a new model is measured against the panel models only. Orig.\ is originality, Range is creative range, and Appr.\ is appropriateness. Cost is the
median generation cost per design in US dollars. $\star$ marks the Pareto frontier of overall score against
cost.}
\label{tab:extended}
\setlength{\tabcolsep}{4pt}
\begin{tabular}{@{}rlrrrrrc@{}}
\toprule
\# & Model & Overall & Orig. & Range & Appr.\ (\%) & Cost (\$) & Pareto \\
\midrule
1 & GPT-6 Astra & 65.5 & 0.646 & 0.472 & 99.1 & 2.564 & $\star$ \\
2 & GPT-6.1 Sol & 63.7 & 0.642 & 0.487 & 98.8 & 0.460 & $\star$ \\
3 & Grok 4.7 & 62.2 & 0.643 & 0.603 & 97.8 & 0.593 &  \\
4 & Gemini 3.8 Flash & 60.6 & 0.637 & 0.628 & 97.5 & 0.123 & $\star$ \\
5 & Qwen3.8 Max & 58.1 & 0.623 & 0.617 & 97.6 & 0.309 &  \\
6 & GPT-5.6 Sol & 51.3 & 0.587 & 0.505 & 98.2 & 0.598 &  \\
7 & Claude Opus 5.5 & 51.0 & 0.557 & 0.574 & 99.2 & 1.557 &  \\
8 & Claude Fable 5.1 & 48.6 & 0.558 & 0.589 & 98.5 & 5.329 &  \\
9 & Muse Spark 1.3 & 47.2 & 0.568 & 0.565 & 97.8 & 0.080 & $\star$ \\
10 & \textit{Hy4 Preview} (high) & 46.2 & 0.580 & 0.605 & 96.8 & 0.086 &  \\
11 & DeepSeek V4.1 Flash & 43.6 & 0.576 & 0.596 & 96.3 & 0.051 & $\star$ \\
12 & Grok 4.6 & 43.4 & 0.591 & 0.625 & 95.5 & 0.100 &  \\
13 & \textit{Pareto 26.10 Preview} (none) & 42.5 & 0.551 & 0.633 & 97.0 & 0.152 &  \\
14 & GLM-5.3 & 41.9 & 0.579 & 0.634 & 95.6 & 0.043 & $\star$ \\
15 & \textit{Mistral Medium 3.5} (high) & 41.5 & 0.675 & 0.643 & 91.0 & 0.080 &  \\
16 & Kimi K3 & 40.4 & 0.547 & 0.615 & 96.8 & 0.570 &  \\
17 & GLM-5.3-Flash & 38.5 & 0.553 & 0.611 & 96.1 & 0.022 & $\star$ \\
18 & \textit{Solar Pro 4} (max) & 36.4 & 0.637 & 0.694 & 91.2 & 0.006 & $\star$ \\
19 & \textit{Seed 2.0 Code} (high) & 34.1 & 0.632 & 0.632 & 91.2 & 0.051 &  \\
20 & MiniMax M3 & 34.0 & 0.587 & 0.602 & 93.4 & 0.024 &  \\
21 & \textit{Laguna S 2.1} (default) & 33.2 & 0.662 & 0.711 & 89.2 & 0.002 & $\star$ \\
22 & \textit{Ling 3.0 Flash VL} (default) & 30.1 & 0.610 & 0.639 & 91.2 & 0.001 & $\star$ \\
23 & \textit{Inkling} (max) & 29.9 & 0.665 & 0.683 & 88.4 & 0.039 &  \\
24 & \textit{Step 3.7 Flash} (high) & 28.1 & 0.627 & 0.646 & 89.9 & 0.015 &  \\
25 & \textit{Solar Mini 4} (default) & 26.0 & 0.649 & 0.694 & 88.1 & 0.004 &  \\
26 & \textit{Mercury 2.5} (high) & 22.5 & 0.730 & 0.631 & 83.8 & 0.001 &  \\
27 & \textit{Nova 2 Lite} (default) & 20.6 & 0.738 & 0.696 & 82.6 & 0.062 &  \\
28 & \textit{Ministral 14B 2512} (none) & 19.0 & 0.723 & 0.622 & 83.3 & 0.001 &  \\
29 & \textit{Relace Search} (none) & 17.7 & 0.738 & 0.617 & 82.3 & 0.013 &  \\
30 & \textit{Command A+} (default) & 10.6 & 0.736 & 0.753 & 80.0 & 0.025 &  \\
31 & \textit{Trinity Large Thinking} (default) & 7.1 & 0.694 & 0.640 & 81.6 & 0.006 &  \\
32 & \textit{Laguna XS 2.1} (default) & 3.3 & 0.701 & 0.696 & 80.1 & 0.001 &  \\
33 & \textit{Seed 1.6 Flash} (default) & -2.1 & 0.810 & 0.704 & 73.7 & 0.001 &  \\
\midrule
-- & Human reference & 92.3 & 0.764 & 0.902 & 98.0 & -- & \\
\bottomrule
\end{tabular}
\end{table}

\section{Example appropriateness criteria}
\label{app:appropriateness}

Table~\ref{tab:appropriateness-example} lists the full criteria set for one design brief from
$\dbench$, a mortgage preapproval form. The brief reads:

\begin{quote}\small
UI Goal: Complete a mortgage preapproval application by entering borrower, property, and loan details, reviewing eligibility, and moving to the next step. \\
Domain: consumer mortgage preapproval application. \\
Brand: Hearthline. \\
Content: borrower Maya Patel, 18 Alder Street, \$420,000 purchase price, \$128,000 annual income, \$84,000 down payment, preapproval status In review.
\end{quote}

\noindent Each criterion is a statement about the finished design that is either true or false.
Criteria marked \emph{applies if} concern elements a design may or may not include; they are true
when the element is absent.

\begin{table}[h]
\centering
\small
\caption{Appropriateness criteria for the mortgage preapproval brief. Critical marks the criteria tagged critical.}
\label{tab:appropriateness-example}
\renewcommand{\arraystretch}{1.15}
\begin{tabular}{@{}l l c p{0.66\linewidth}@{}}
\toprule
ID & Category & Critical & Criterion \\
\midrule
C01 & Scope & \checkmark & The page is a mortgage preapproval application screen where the applicant enters details, not a marketing landing page, rate-comparison page, or sign-in page. \\
C02 & Scope &  & The page shows that the application is a multi-step flow and marks which step is current. \\
C03 & Content &  & The brand name Hearthline appears as the app's identity. \\
C04 & Content & \checkmark & The borrower is shown as Maya Patel. \\
C05 & Content & \checkmark & The property address is shown as 18 Alder Street. \\
C06 & Content & \checkmark & The purchase price is shown as exactly \$420,000. \\
C07 & Content & \checkmark & The borrower's annual income is shown as exactly \$128,000. \\
C08 & Content & \checkmark & The down payment is shown as exactly \$84,000. \\
C09 & Content & \checkmark & A preapproval status is shown with the value 'In review'. \\
C10 & Actions & \checkmark & Input fields for borrower details (at least name and annual income) are shown. \\
C11 & Actions & \checkmark & Input fields for property details (at least address and purchase price) are shown. \\
C12 & Actions & \checkmark & Input fields for loan details (at least down payment, plus loan amount, loan type, or term) are shown. \\
C13 & Actions & \checkmark & A control to continue to the next step of the application is shown. \\
C14 & Correctness & \checkmark & Any loan amount shown equals \$336,000 (\$420,000 purchase price minus \$84,000 down payment). \emph{(Applies if a loan amount is shown.)} \\
C15 & Correctness &  & Any down payment percentage shown is 20\%, and any loan-to-value ratio shown is 80\%. \emph{(Applies if a down payment percentage or LTV is shown.)} \\
C16 & Correctness &  & Any value that appears more than once (purchase price, income, down payment, address, borrower name) shows the same value everywhere it appears. \emph{(Applies if a value appears more than once.)} \\
C17 & Domain &  & The eligibility review names at least one specific factor the preapproval depends on (e.g. income, down payment/LTV, debt-to-income, or credit) with its value or result. \\
C18 & Goal &  & An eligibility review section or summary is shown together with the entered borrower, property, and loan details on this screen. \\
\bottomrule
\end{tabular}
\end{table}

\end{document}